 \documentstyle {article}
\newtheorem{theorem}{Theorem}[section]
\newtheorem{proposition}{Proposition}[section]
\newtheorem{definition}{Definition}[section]
 \font\frakfont=euphm10
\def\frak#1{\hbox{\frakfont #1}}

\begin{document}
 \title {Poisson Lie Groups, Quantum Duality Principle,
  and the Quantum Double}
 \author {M.A.Semenov-Tian-Shansky}
 %
  \thanks{The author is grateful to the Organizing Committee of the
  Mt.Holyoke Conference and to Universit\'{e} de Bourgogne, Dijon,
  France, for financial subsistence}
 \maketitle
 \begin{abstract}
 The Heisenberg double of a Hopf algebra may be regarded as a quantum
 analogue of the cotangent bundle of a Lie group. Quantum duality
 theory describes  relations between a Hopf algebra, its dual, and their
 Heisenberg double in a way which extends both the theory of coadjoint
 orbits and the classical Fourier transform. We also describe the
 twisted Heisenberg double which is relevant for the study of
 nontrivial deformations of the quantized universal enveloping algebras.
\end{abstract}
 \section{Introduction}.

 The standard way to describe quantum deformations of simple finite
 dimensional or affine Lie algebras is by means of generators  and
 relations (generalizing the classical Chevalley - Serre relations)
  \cite{dr,j}. A dual approach, due to Faddeev, Reshetikhin, and
 Takhtajan \cite{frt}, \cite{rtf}, is
 to construct quantum universal enveloping algebras as deformations of
 coordinate rings of functions on Lie groups (regarded as affine
 algebraic groups). Of course, construction of a quantum deformation of
 the algebra $Fun(G)$ was one of the first results of the quantum group
 theory and is, in
fact, a direct generalization of the Baxter
 commutation relation $R T_{1}T_{2} ~~=~~T_{2}T_{1} R$. A nontrivial fact,
 first observed by Faddeev, Reshetikhin, and Takhtajan, is that the dual
 algebra $ Fun_{q}(G)^{*}$ may also be regarded as a deformation of a
 function algebra on a Lie group (namely, on the dual group $G^{*}$, see
 below)\footnote{In a disguised form this fact was also mentioned in
 Drinfeld's report \cite{dr}(`Equivalence of the category of
 QFSH-agebras and the category of QUE-algebras')} . More generally, the
  FRT construction is related to the {\em
 quantum duality principle} which we are now going to state.

   Observe, first of all, that in the semiclassical approximation quantum
   deformations of function algebras are determined by Poisson brackets
   on Lie groups. The class of Poisson brackets related to deformations
   of algebras $Fun(G)$ in the category of Hopf algebras is described by
   the following well-known axiom \cite{dr1}.

   \begin{definition}
   A Poisson bracket on a Lie group $G$ defines on $G$ the structure of
   a Poisson Lie group if multiplication
$$m:G \times G \rightarrow G$$
is a Poisson map
\end {definition}
  A Poisson bracket satisfying this axiom is degenerate and, in
  particular, is identically zero at the unit element of the group.
  Linearizing it at this point defines the structure of a Lie algebra in
  the space $T^{*}_{e}G \simeq \frak{g}^{*}$. The pair
  $(\frak{g},\frak{g}^{*})$ is called the {\em tangent Lie bialgebra} of
  $G$.
  Lie brackets in $\frak{g}$ and $\frak{g}^{*}$ satisfy the following
  compatibility condition:

   {\em Let }$\varphi:\frak{g}\rightarrow \frak{g}\wedge \frak{g}$ {\em be
 the dual  of the commutator map } $[,]_{*}: \frak{g}^{*}\wedge
  \frak{g}^{*}\rightarrow \frak{g}^{*}$.{\em Then }$\varphi $ {\em is a
  1-cocycle on} $  \frak{g}$ {\em (with respect to the natural action
  of } $\frak{g}$ {\em on} $  \frak{g}\wedge\frak{g})$.

     Let $c_{ij}^{k}, f^{ab}_{c}$ be the structure constants of
     $\frak{g}, \frak{g}^{*}$ with respect to the dual bases $\{e_{i}\},
     \{e^{i}\}$ in $\frak{g},\frak{g}^{*}$. The compatibility condition
     means that
     $$c_{ab}^{s} f^{ik}_{s} ~-~ c_{as}^{i} f^{sk}_{b} ~+~ c_{as}^{k}
     f^{si}_{b} ~-~ c_{bs}^{k} f^{si}_{a} ~+~ c_{bs}^{i} f^{sk}_{a} ~~=
     ~~0.$$
     This condition is symmetric with respect to exchange of $c$ and
     $f$. Thus if $(\frak{g},\frak{g}^{*})$ is a Lie bialgebra, then
     $(\frak{g}^{*}, \frak{g})$ is also a Lie bialgebra. Let $G^{*}$ be a
     (connected simply connected) Lie group which corresponds to
     $\frak{g}^{*}$. Since the correspondence between Poisson Lie groups
     and Lie bialgebras is functorial, $G^{*}$ is also a Poisson Lie
     group (called the dual of $G$).

       Passing to quantization, we may (at least if there are no
       obstructions, see \cite{dr}) construct two Hopf algebras
       $Fun_{q}(G),~
       Fun_{q}(G^{*})$  which correspond to the Poisson---Hopf algebras
       $Fun(G), Fun(G^{*})$.The {\em quantum duality principle} then
       asserts that these algebras are dual to each other as Hopf
       algebras. More precisely, let $h$ be the deformation parameter
       (for simplicity we chose $h$ to be the same for both algebras).
       There exists a nondegenerate bilinear pairing
       $$Fun_{q}(G)~\otimes~ Fun_{q}(G^{*}) \rightarrow
       \bf{C}\mbox{$[[h]]$}$$
       which sets the algebras $Fun_{q}(G), Fun_{q}(G^{*})$ into
       duality as
       Hopf algebras. Hence, in particular, we have, up to an
       appropriate completion,
       $$Fun_{q}(G^{*})\simeq U_{q}(\frak{g}).$$
       In the dual way, we have also
       $$Fun_{q}(G) \simeq U_{q}(\frak{g}^{*}).$$

       As a simple example, let us consider {\em trivial Lie bialgebra}.
       Let
       $\frak{g}$ be an arbitrary Lie algebra, $\frak{g}^{*}$ its dual
       equipped with the zero  Lie bracket. In this case the Poisson
       bracket on $G$ is trivial. The dual group of $G$ is the additive
       group of the space $\frak{g}^{*}$ equipped with the Lie---Poisson
       bracket \cite{ber}, \cite{wein} of $\frak{g}$. In the present case
       the algebra  $Fun(G)$
       does not deform at all (since the germ of a deformation defined
       by the Poisson bracket is identically zero). The deformation of
       $Fun(\frak{g}^{*})$ may be identified with the universal
       enveloping algebra $U(\frak{g})$ \cite{ber}, a function $\psi\in
       Fun(\frak{g}^{*})$ being regarded as the symbol of a left
       invariant (pseudo)differential operator on $G$. The pairing
       $Fun(G)~\otimes~ Fun(\frak{g}^{*})\rightarrow
       \bf{C}\mbox{$[[h]]$}$ is given
       by
       \begin{equation}
       <\varphi,\psi>~~=~~\int_{G \times\frak{g}^{*}}\varphi(x)\psi(p)
       \exp i/h<p,\log x> dx dp
       \label{dxdp}
        \end{equation}
       The integration measure $ dx dp $ in (\ref{dxdp}) is of course the
       Liouville measure on $G\times \frak{g}^{*}\simeq T^{*}G.$ The
       emergence of  $T^{*}G$ in this context is not accidental. The
       pairing (\ref{dxdp}) canonically generates an action
       $$Fun_{q}(\frak{g}^{*})~\otimes~ Fun(G)\rightarrow Fun(G):
        \psi\otimes\varphi\rightarrow <id\otimes\psi,\Delta\varphi>$$
        which is the usual action of $U(\frak{g})$ on $Fun(G)$ by
        left-invariant derivations. Let $\cal H$ be the associative
        algebra generated by $U(\frak{g})$ and $Fun(G)$ regarded as
        differential operators and multiplication operators in $Fun(G)$,
        respectively. Then $\cal H$ is a quantum deformation of the
        Poisson algebra of functions on $T^{*}G$ with the canonical
        Poisson bracket. The algebra $\cal H$ arises along with its
        irreducible (Schroedinger) representation. As we shall see
        below, $\cal H$ is a special case of the {\em Heisenberg
        double}.

          Thus even for trivial Lie bialgebras the quantum duality
          principle is quite meaningful: it includes, e.g., the usual
          Fourier duality. In general case, quantum duality principle
          may also be regarded as a generalization of the Fourier
          transform (cf. Section 4).

          For semisimple Lie algebras quantum deformations of Poisson
          algebras $Fun(G)$, $ Fun(G^{*})$ may easily be constructed
          if we know the corresponding  quantum R-matrices.
          This allows also to check the duality principle. The need to
          know  the quantum R-matrices in advance is a certain drawback
          of this approach, as compared to the standard definition of
          Drinfeld and Jimbo. On the other hand, the  FRT construction
          alllows to define, along with the well-known algebras, such as
          $Fun_{q}(G)$ and $Fun_{q}(G^{*})\simeq U_{q}(\frak {g})$, a
          large family of their relatives. The two simplest algebras in
          this family are the {\em Heisenberg double} and the {\em
          Drinfeld double}. Both are naturally described as
          quantizations of some geometrically defined Poisson algebras.
          (An elementary algebraic definition in these cases is also
          possible.\footnote{The geometric construction of the
          Heisenberg double was proposed almost simultaneously by the
          author and by Alekseev and Faddeev \cite{af}.The author has
          also pointed out its direct algebraic definition. Recently,
           an algebraic construction of a family of operator algebras
            which includes the Heisenberg double and the  Drinfeld double,
             was proposed by S.P.Novikov \cite{nov}})
            We have seen already, on an elementary example,
          that the Heisenberg double is the quantum analogue of the
          algebra $Fun(T^{*}G)$. Recall that connections between the
          canonical Poisson bracket on $T^{*}G$ and the Lie---Poisson
          bracket on $\frak{g}^{*}$ form the basis of the standard
          geometric construction of irreducible representations of $G$
          (the so called `orbits method'). Mutual relations between the
          algebras $Fun_{q}(G), Fun_{q}(G^{*})$ and their Heisenberg
          double  form the precise analogue of the orbits method ( the
          author hopes to describe this subject more thoroughly in a
          separate paper).

            In many cases the Heisenberg double admits non-trivial
            deformations associated with outer automorphisms of the
            underlying algebras. In this way we are led to the notion of
            the {\em twisted double}. This allows notably to define  the
            twisted algebra $U_{q}(\frak{g})_{\tau}^{\otimes N} $ which
            may be regarded as a lattice version of a current algebra,
            the twisting carefully reproducing the effects of central
            extension \cite{afs}.\footnote{Recently Fock and Rosly
            introduced a
            still more general class of algebras which are associated
            with arbitrary graphs \cite{fock}.} The author does not
            know any
            direct elementary definition of the twisted double or of
            the related algebras, so here the geometric approach seems
            indispensable.

            For general semisimple Lie algebras the logics of our
            construction is as follows:
            \begin{enumerate}
            \item We use the Drinfeld---Jimbo construction to define
            quantum
            deformations of the universal enveloping algebras
            \item The universal R-matrix theory \cite{kr,ls} and the
            representation theory for $U_{q}(\frak{g})$ yield explicit
            formulae for quantum R-matrices and for the FRT generators
            of the algebras $Fun_{q}(G), Fun_{q}(G^{*})$ in terms of the
            Drinfeld---Jimbo generators.
            \item Finally, we may describe all related algebras (the
            Heisenberg  double, the twisted double and their
            subalgebras).
            \end{enumerate}

            In the present article we shall leave aside the theory of
            universal R-matrices and admit (in the spirit of \cite{frt})
            that
            all necessary R-matrices are known in advance.

 {\bf Acknowledgement.} The author had numerous fruitful discussions on
  the
matters discussed in this paper with A.Alekseev, L.D.Faddeev and
N.Reshetikhin.

\section {   Quasiclassical Duality Theory}
\setcounter{equation}{0}
     Let $(\frak{g},\frak{g}^{*})$ be  Lie bialgebra. Put
     $\frak{d}~~=~~\frak{g}\oplus\frak{g}^{*}.$ There is a unique
     structure
     of a Lie algebra on $\frak{d}$ such that
     \begin{enumerate}
     \item[(i)] $\frak{g}, \frak{g}^{*}\subset\frak{d} $ are Lie
     subalgebras.
     \item[(ii)] The canonical bilinear form on $\frak{d}$ given by
     \begin{equation}
     <(X_{1},f_{1}),~(X_{2},f_{2})>~~=~~f_{1}(X_{2})~+~f_{2}(X_{1})
     \label{inprod}
     \end{equation}
     is $ad-\frak{d}$-invariant.
     \end{enumerate}
     Let $P_{\frak{g}}, P_{\frak g^{*}}$ be the projection operators onto
     $\frak{g},\frak g^{*}\subset\frak{d}$ parallel to the complementary
     subalgebra. The linear operator
     \begin{equation}
     r_{\frak{d}}~~=~~P_{\frak{g}}~-~P_{\frak{g}^{*}}
     \label{rd}
     \end{equation}
     is skew-symmetric with respect to (\ref{inprod}) and may be
     identified with
     an element of $\wedge^{2} \frak{d}$. The formula
     \begin{equation}
     [X,Y]_{*}~~=~~\frac{1}{2}(~[r_{\frak{d}}X,~Y]~+~[X,~r_{\frak{d}}Y]~)
     \label{rbr}
     \end{equation}
     defines in the space $\frak{d}^{*}\simeq \frak{d}$ a Lie bracket
     which makes $(\frak{d}, \frak{d}^{*})$ a Lie bialgebra. As a linear
     space again $\frak{d}^{*}~~=~~\frak{g}\oplus\frak{g}^{*}$ but this
     time
     $\frak{g}, \frak{g}^{*} $  are complementary ideals in $\frak{d}^{*}$;
     moreover, $\frak{d}^{*}/\frak{g}^{*}$ is isomorphic to $\frak{g}$ and
     $\frak{d}^{*}/\frak{g}$ is {\em anti-isomorphic} to $\frak{g}^{*}$.
     The Lie
     bialgebra $(\frak{d}, \frak{d}^{*})$  is called the double of $(\frak
     g, \frak{g}^{*})$. Clearly, the dual Lie bialgebras $(\frak{g},\frak
     g^{*})$ and $(\frak{g}^{*},\frak{g})$ have a common double.

     Let $D$ be the (connected simply connected) Lie group with the Lie
     algebra $\frak{d}$. One can define on $D$ several important Poisson
     structures. The two simplest ones are defined as follows.

       For $\varphi\in C^{\infty}(D)$ let $X_{\varphi},
       X'_{\varphi}\in \frak{d}$ be its left and right gradients
       defined by the formulae
       \begin{eqnarray}
       <X_{\varphi}(x),\xi>~=~\frac{d}{dt}_{t=0}\varphi(e^{t\xi}x),
       \nonumber\\
       <X'_{\varphi}(x),\xi>~ =~\frac{d}{dt}_{t=0}\varphi(xe^{t\xi}),~~
       \xi\in\frak{d}.
       \label{grad}
       \end{eqnarray}
       Put
       \begin{equation}
       \{\varphi,\psi\}_{\pm}~~=~~\frac{1}{2}<r_{\frak{d}}X_{\varphi},
       X_{\psi}>~\pm~
       \frac{1}{2}<r_{\frak{d}}X^{\prime}_{\varphi},X^{\prime}_{\psi}>.
       \label{pbr}
       \end{equation}
       The bracket $\{,\}_{-}$ equips $D$ with the structure of a
       Poisson Lie group; its tangent Lie bialgebra is precisely $(\frak
       d, \frak{d}^{*})$. The bracket $\{,\}_{+}$ is non-degenerate (at
       least on an open dense subset in $D$)  and defines (almost
       everywhere) on $D$ a symplectic structure. If $(\frak{g},\frak
       g^{*})$ is a trivial Lie bialgebra, the group $D~=~G\times\frak
       g^{*}$ is the semi-direct product of $G$ and the additive group
       of $\frak{g}^{*}$. Thus $D$ may be identified with $T^{*}G$. It is
       easy to check that $\{,\}_{+}$  coincides in this case with the
       canonical  Poisson bracket on $T^{*}G$. The bracket  $\{,\}_{-}$
       is highly degenerate: it is the direct product of the
       Lie---Poisson bracket on $\frak{g}^{*}$ and the trivial bracket on
       $G$.

       In general, $\{,\}_{+}$ is also an analogue of the canonical
       Poisson bracket on the cotangent bundle. In order to describe its
       relations with the Poisson brackets on $G$ and on $G^{*}$,  let us
       first recall some simple facts on Poisson reduction \cite{rims,lu}.

         Let $M$ be a Poisson manyfold. An action $H\times M\rightarrow
         M$ is called {\em admissible} if the space of $H$-invariant
         functions
         is  a Lie subalgebra of the Poisson algebra $Fun(M)$. Admissible
         actions may {\em not} preserve Poisson brackets on $M$. If $H
         \times M\rightarrow M$ is an admissible action and the quotient
         space $M/H$ is smooth, we may identify the algebras $Fun(M/H)$
         and $Fun(M)^{H}$. Hence there exists a Poisson structure on $M/H$
         such that the canonical projection $\pi:M\rightarrow M/H$ is a
         Poisson map. The space $M/H$ is called the {\em reduced Poisson
         manyfold.} Even if $M$ is symplectic, the reduced Poisson
         bracket on $
         M/H$ is usually degenerate. The difficult part of
         reduction is the description of its symplectic leaves. To solve
         this problem we need the notion of {\em dual pairs} \cite{wein}.

          Assume that there are {\em two} transformation groups
          $H,H^{\prime}$ acting on $M$. Admisssible actions $H\times
          M\rightarrow M, H^{\prime}\times M\rightarrow M$ are said  to
          be {\em dual} to each other if the subalgebras of invariants
          $Fun(M)^{H},~ Fun(M)^{H^{\prime}}$ are the centralizers of each
          other in the Poisson algebra $Fun(M)$. Assume that $M$ is
          symplectic and $H,H^{\prime}$ are dual transformation groups
          of $M$. In that case symplectic leaves in $M/H$  are the
          connected components of the sets $\pi (\mbox{$\pi'$}^{ -1}(x)),
          x\in M/H'$; in a similar way, symplectic leaves in
          $M/H'$  are the connected components of the sets
          $\pi'(\pi^{-1}(y)), ~y\in M/H$.

          \begin{theorem}
          Let $G,G^{*}$  be the dual Poisson---Lie groups, $D_{+}$ their
          double equipped with the Poisson bracket $\{,\}_{+}$
          (\ref{pbr}). Then
          \begin{enumerate}
          \item[(i)] Natural actions of $G(G^{*})$ on $D$  by left and
          right
          translations are admissible.
          \item[(ii)] The actions of $G (G^{*})$  on $D_{+}$  by left
          and right
          translations form a dual pair.
          \end{enumerate}
         \label{th2.1}
          \end{theorem}
            Elements in $D$  which admit a unique factorization
            $$ x~=~g\cdot g^{*}~=~\tilde{g}^{*}\cdot \tilde{g}~,~~~~g,
            \tilde{g}\in
            G~,~~~~
            g^{*},\tilde{g}^{*}\in G^{*},$$
            form an open dence subset in $D$. Thus $G$ may be identified
            with an open dense subset in $D/G^{*}$, or in $G^{*}\backslash
            D$, and $G^{*}$ with an open dense subset in $D/G$, or in
            $G\backslash D$.

            \begin{theorem}
            \begin{enumerate}
            \item[(i)] $G\subset D/G^{*}$  is a Poisson submanifold; the
            induced Poisson structure on $G$ is anti-isomorphic to the
            original one.
            \item[(ii)] In a similar way, $G^{*}$  is a Poisson
            submanyfold in
            $ D/G$; the induced Poisson bracket on $G^{*}$ coincides
            with the original one.\footnote{Sign difference in (i), (ii)
            is due to the minus sign in (\ref{rd}).}
            \end{enumerate}
            \label{th2.2}
            \end{theorem}
            For trivial Lie bialgebras $D~=~T^{*}G,D/G\simeq \frak
            g^{*}$, and Theorem \ref{th2.2} amounts to the well known
            connection
            between the canonical Poisson bracket on $T^{*}G$ and the
            Lie---Poisson bracket on $\frak{g}^{*}$. Theorem \ref{th2.1}
            then
            asserts that the Hamiltonians of left and right translations
            on $T^{*}G$ are in involution with respect to the canonical
            Poisson bracket.This result plays the key role in the
            classical `orbits method'.

               We shall be mainly concerned with a special class of Lie
               bialgebras, the so called {\em factorizable Lie
               bialgebras}. Let $\frak{g}$  be a semisimple Lie algebra
               equipped with a fixed nondegenerate invariant inner
               product. We  use it to fix an isomorphism of the dual
               space $\frak{g} ^{*}$ with $\frak{g}$. The bialgebra
               structure on $\frak{g}$ is defined by the cobracket
               $$\phi (X)~=~-~\frac{1}{2}[r,X\otimes 1+1\otimes X],$$
               where $r\in \wedge^{2}\frak{g}$  is a classical r-matrix.
               Since we identified $\frak{g}^{*}$ with $\frak{g}$, we may
               regard $r$ as a skew symmetric linear operator in $\frak
               g$. Assume that $r$ satisfies the {\em modified classical
               Yang---Baxter identity}
               \begin{equation}
               [rX,rY]~~=~~r(~[rX,~Y]~+~[X,~rY]~)~-~[X,Y].
               \label{cybe}
               \end{equation}
               The Lie bracket on $\frak{g}^{*}\simeq\frak{g}$ which
               corresponds to (\ref{cybe}) is given by
               \begin{equation}
               [X,Y]_{*}~=~\frac{1}{2}([rX,Y]~+~[X,rY])
               \label{rb}
               \end{equation}
               and by virtue of (\ref{cybe}) satisfies the Jacobi
               identity. Put
               $ r_{\pm}~~=~~\frac{1}{2}(r\pm id)$. Then (\ref{cybe})
               implies that $r_{\pm}$
               regarded as a mapping from $\frak{g}^{*}$ into  $\frak{g}$
               is a Lie algebra  homomorphism. Put $\frak{d} ~~=~~\frak
               g\oplus\frak{g}$ (direct sum of two copies).The mapping
               $$\frak{g}^{*}\rightarrow \frak{d}~~~:X\mapsto
               (X_{+},~X_{-}),~~~X_{\pm}~=~r_{\pm}X,$$
               is a Lie algebra embedding. Thus we may identify $\frak
               g^{*}$ with a Lie subalgebra in $\frak{d}$. Let $\frak
               g^{\delta}\subset \frak{d}$ be the diagonal subalgebra.
               Equip $\frak{d}$ with the inner product
               \begin{equation}
               \ll(X,X^{\prime}), (Y,Y^{\prime})\gg~~=~~
                <X,Y>-<X^{\prime},Y^{\prime}>.
               \label{prod}
               \end{equation}

                \begin{proposition}
                \begin{enumerate}
                \item[(i)] As a  linear space
                $$\frak{d}~~=~~\frak{g}^{\delta}\oplus \frak{g}^{*}.$$

                \item[(ii)] Let $P_{\frak{g}}, ~P_{\frak{g}^{*}}$ be the
                projection operators onto $\frak{g}^{\delta},~\frak
                g^{*}\subset\frak{d}$ parallel to the complementary
                 subalgebra,$r_{\frak{d}}~~
                 =~~P_{\frak{g}}-P_{\frak{g}^{*}}$.
                Then $r_{\frak{g}}$ is skew with respect to the inner
                product (\ref{prod}) and satisfies (\ref{cybe}).
                Hence it defines on
                $\frak{d}$ the structure of a Lie bialgebra.
                \item[(iii)] $(\frak{d},\frak{d}^{*})$ is canonically
                isomorphic
                to the double of $(\frak{g}, \frak{g}^{*})$.
                \end{enumerate}
                \label{pr1}
                \end{proposition}
                The explicit expression for $r_{\frak{d}}\in End(\frak
                g\oplus\frak{g})$  in terms of the original r-matrix is
                given by
                \begin{equation}
                r_{\frak{d}}~~~=~~~\left(
                \begin{array}{ll}
                r & -2r_{+}\\
                2r_{-} & -r
                \end{array}
                \right).
                \label{rdouble}
                \end{equation}
A Lie bialgebra $(\frak{g},\frak{g}^{*})$  with the properties as above is
called a {\em factorizable Lie bialgebra}. Thus the double of a
factorizable Lie bialgebra is isomorphic (as  a Lie algebra) to the
square of $\frak{g}$.

Now let $G$ be a linear algebraic group with the Lie algebra $\frak{g}$.
Put $D~~=~~G\times G$. Embedding $\frak{g}^{*}\hookrightarrow \frak{d}$
may
be extended to a homomorphism $G^{*}\hookrightarrow D$; we shall
identify $G^{*}$ with the corresponding subgroup in $D$. Almost all
elements $(x,y)\in D$ admit a representation
\begin{equation}
(x,y)~~=~~(L^{+},L^{-})\cdot (T,T)^{-1},
\label{fact}
\end{equation}
where $(L^{+},L^{-})\in G^{*},~~(T,T)\in G^{\delta}\subset D$.

Let $(\rho,V)$ be an exact matrix representation of $G$. The algebra
$Fun(G)$ is generated by matrix coefficients $\rho (x)_{ij},~\rho
(y)_{ij}$. Matrices $L^{\pm},T$ may be regarded as (almost everywhere
regular) functions of $x,y$. Hence the matrix coefficients $\rho
(L^{\pm})_{ij},~~\rho (T)_{ij}$ give another system of generators of the
(suitably enlarged) algebra $Fun(D)$. Functions $\rho(L^{\pm}_{ij},
\rho(T)_{ij}$  are rational functions on $D$ with singularities at those
points $(x,y)\in G$ for which factorization (\ref{fact}) does not exist.

It is convenient to define the Poisson structure on $D$ in terms of the
generators of $Fun(D)$. We use the standard tensor notation to suppress
matrix indices. Thus we write $T_{1}~=~T\otimes id, ~~T_{2}~=~id\otimes
T$, etc. The Poisson bracket $\{T_{1}^{V},T_{2}^{W}\}$ is,
by definition, a
matrix in $End (V\otimes W)$ whose matrix coefficients are the Poisson
brackets $\{\rho_{V}(T)_{ij}, \rho_{W}(T)_{kl}\}$. The superscripts
$V,W$ will sometimes be omitted. We shall need explicit formulae for the
Poisson brackets of two systems of generators of $Fun(D)$. We have

\begin{eqnarray}
\{x_{1},x_{2}\}_{\pm}~=~\frac{1}{2}~(
r_{VW}~x_{1}x_{2}~\pm~x_{1}x_{2}~r_{VW}~),\nonumber\\
\{y_{1},y_{2}\}_{\pm}~=~\frac{1}{2}~(
r_{VW}~y_{1}y_{2}~\pm~y_{1}y_{2}~r_{VW}~),\nonumber\\
\{y_{1},x_{2}\}_{\pm}~=~r^{+}_{VW}~y_{1}y_{2}~
\pm~y_{1}x_{2}~r^{+}_{VW},~~~\label{xybr}
\end{eqnarray}

\begin{eqnarray}
\{T_{1},T_{2}\}_{\pm}~=~\frac{1}{2}~[r_{VW},T_{1}T_{2}],\nonumber\\
\{L^{\pm}_{1},L^{\pm}_{2}\}_{\pm}~=~
\frac{1}{2}~[r_{VW},L^{\pm}_{1}L^{\pm}_{2}],\nonumber\\
\{L^{+}_{1},L^{-}_{2}\}_{\pm}~=~[r^{+}_{VW},L^{+}_{1}L^{-}_{2}],
\nonumber\\
\{L^{\pm}_{1},T_{2}\}_{+}~=~L^{\pm}_{1}T_{2}~r^{\pm}_{VW},\nonumber\\
\{L^{\pm}_{1},T_{2}\}_{-}~=~0.
\label{doublebr}
\end{eqnarray}

Here $r_{VW}~~=~~(\rho_{V}~\otimes~\rho_{W})r\in End(V\otimes W)$ and
$r$ is
regarded as an element of $\wedge ^{2}\frak{g}$.  Note that Poisson
brackets $\{,\}_{\pm}$ in terms of the generators $L^{\pm},T$ differ
only by the expression for $\{L^{\pm}_{1},T_{2}\}$. Recall that $D_{-}$
is  a Poisson Lie group. Formulae (\ref{doublebr}) show that as
a Poisson manyfold
(though of course not as a group) $D_{-}$ is a direct product of its
subgroups $G,G^{*}$. The brackets $\{,\}_{+}$ are non-degenerate.

Since $D_{-}$  is a Poisson Lie group, the diagonal map
$$\Delta:Fun(D_{-})\rightarrow Fun(D_{-}\times D_{-})$$
is a homomorphism of Poisson algebras. This map may easily be described
in terms of the generators $x,y$:
$$\Delta x~ =~x~\dot{\otimes}~x,~~\Delta~y~=~ y~\dot{\otimes}~y,$$
or, in a more accurate notation,
$$\Delta(\rho(x)_{ij})~=~\sum _{k} \rho(x)_{ik}\otimes\rho(x)_{kj},$$
and similarly for $\Delta (\rho(y))$. It is easy to check that
multiplication in $D$ induces a Poisson mapping $D_{-}\times
D_{+}\rightarrow D_{+}$. Hence the diagonal map also defines on
$Fun(D_{+})$ the structure of a left $Fun(D_{-})$-comodule. In the
obvious notation we may write
$$C(x_{+})~=~x_{-}~\dot{\otimes}~x_{+},
{}~~C(y_{+})~=~y_{-}~\dot{\otimes}~y_{+}.$$

Another group which is also important for the duality theory is the {\em
dual double} $D^{*}$. By definition, $D^{*}$ is the Poisson---Lie group
which corresponds to the Lie bialgebra $(\frak{d}^{*},\frak{d})$. Since
$\frak{d}^{*}\simeq \frak{g}\oplus\frak{g}^{*}$ is a direct sum of Lie
algebras, $D^{*}~=~G\times G^{*}$ as a Lie group. However, the Poisson
structure on $D^{*}$ does {\em not} split. The algebra $Fun(D^{*})$ is
generated by matrix coefficients $\rho (L^{\pm},\rho (T)$. Let us denote
these generators by $^{*}L^{\pm},~^{*}T$ in order to distinguish them
from the generators (\ref{fact}). We have

\begin{eqnarray}
\{^{*}T_{1}~,~^{*}T_{2}\}~=~\frac{1}{2}~[r~,~^{*}T_{1}~^{*}T_{2}],
{}~~~\nonumber\\
\{^{*}L^{\pm}_{1}~,~^{*}L^{\pm}_{2}\}~=~
\frac{1}{2}~[r~,~^{*}L^{\pm}_{1}~^{*}L^{\pm}_{2}],~\nonumber\\
\{^{*}L^{+}_{1}~,~^{*}L^{-}_{2}\}~=~
[r~,~^{*}L^{+}~^{*}L^{-}_{2}\},\nonumber\\
\{^{*}L^{\pm}_{1}~,~^{*}T_{2}\}~=~[r^{\pm}~,~^{*}L^{\pm}_{1}~^{*}T_{2}].
\label{double*}
\end{eqnarray}

 The coproduct in $Fun(D^{*})$ is given by
 $$\Delta^{*}T~~=~~^{*}T~\dot{\otimes}~^{*}T,~~\Delta~^{*}L^{\pm} =
 ^{*}L^{\pm} \dot{\otimes
 }^{*}L^{\pm}.$$
 Let us return to the study of the bracket $\{,\}_{+}$ on $D$ and
 discuss the relations between various systems of generators of
 $Fun(D_{+})$. Observe first of all that due to factorization formula
  (\ref{fact}) we may regard $L^{\pm}$ as generators of the algebra
 $Fun(D/G^{\delta}$ and $T$ as generators of the algebra
 $Fun(G^{*}\backslash D)$ .One can also factorize $(x,y)$ in the
 opposite order:
 \begin{equation}
 (x,y)~~=~~(\hat{T},\hat{T})\cdot (\hat{L}^{+},\hat{L}^{-})^{-1}
 \label{oppfact}
 \end{equation}
  Clearly, the matrix coefficients of $\hat{T}$ generate the algebra
  $Fun(G^{\delta}\backslash D)$. As we know, canonical projections
 \begin{equation}
  D/G \stackrel{\pi}{\longleftarrow}D\stackrel{\prime{\pi}}
  {\longrightarrow}
 G\backslash D,~~~~~~
 D/G^{*}\stackrel{p}{\longleftarrow}D\stackrel{p'}
 {\longrightarrow}G^{*}\backslash
 D
 \label{diagr}
 \end{equation}
  form dual pairs. Hence the generators $\hat{L}^{\pm},\hat{T}$ satisfy
  the Poisson bracket relations (\ref{doublebr}) and, moreover,
  $$\{L^{\pm}_{1}~,~\hat{L}^{\pm}_{2}\}~=~\{T_{1}~,~\hat{T}_{2}\}~=~0.$$
  Decompositions (\ref{fact}, \ref{oppfact}) are analogous to the
  definition
  of left and right momenta for a rigid body \cite{arn}, or to chiral
  decompositions in Conformal Field Theory (the last analogy is
  discussed in \cite{af}).

 The quotient spaces $D/G^{\delta},~G^{\delta}\backslash D$ may be
 canonically identified with the group G itself. The projection maps
 $\pi,\hat{\pi}$ are given by
 $$\pi (x,y)~~=~~xy^{-1},~~~~~~\hat{\pi}(x,y)~~=~~y^{-1}x.$$
 This allows to introduce another set of generators for the algebras
 $Fun(D/G)$, $Fun(G\backslash D)$. Put
 $L_{V}~~=~~\rho_{V}(xy^{-1}),\hat{L}_{V}~~=~~\rho_{V}(y^{-1}x)$ (here, as
 above, $(\rho,V)$ is an exact linear representation of $G$ ). Clearly,
we  have
 $$L_{V}~~=~~L^{+}_{V}(L^{-}_{V})^{-1},\hat{L}_{V}~~=~~
 (\hat{L}^{+}_{V})^{-1}\hat{L}^{-}_{V}.$$
 Note that matrix coefficients of $L_{V},\hat{L}_{V}$ are everywhere
 regular functions on $D$. The Poisson brackets for these generators are
 given by
\begin{equation}
\{L_{1}~,L_{2}\}~~=~~L_{1}r_{+}L_{2}~+~L_{2}r_{-}L_{1}~-
{}~\frac{1}{2}~L_{1}L_{2}r~-~\frac{1}{2}~rL_{1}L_{2},
\label{g*br}
\end{equation}

\begin{equation}
\{L_{1}~,~T_{2}\}~~=~~L_{1}T_{2}r_{-}~-~T_{2}r_{+}L_{1}.
\label{LTbr}
\end{equation}

According to the general theory, symplectic leaves in $D/G\simeq G$ are
the connected components of the sets $\pi (\mbox{$\pi'$}^{-1}(x)),x\in
G\backslash D$. In the present case
$\pi(x,y)~~=~~xy^{-1},~~~\pi'(x,y)~~=~~y^{-1}x$. Thus symplectic
leaves in $G$
are simply the conjugacy classes; moreover,the action of $G$ on itself
by conjugations is a Poisson mapping
\begin{equation}
G\times G\rightarrow G:(x,L)\mapsto xLx^{-1}.
\label{conj}
\end{equation}
Here the bracket $\{L_{1},L_{2}\}$ is given by (\ref{g*br}), and the
bracket for the matrix coefficients of $x$ is given by the standard
formula
\begin{equation}
\{x_{1}~,~x_{2}\}~=~\frac{1}{2}~[r~,~x_{1}x_{2}].
\label{sklbr}
\end{equation}
By duality, we get a morphism
\begin{equation}
Fun(D/G)\rightarrow Fun(G)~\otimes~ Fun(D/G):~~L\mapsto
x_{1}L_{2}x_{1}^{-1}.
\label{conj*}
\end{equation}
The action (\ref{conj}) and the dual coaction (\ref{conj*})
is an example of
the so called {\em dressing transformations} \cite{rims}.

Let us finally note that the Casimir functions which form the center of
the Poisson algebra $Fun(D/G)$ are precisely central functions on $G$.
The generators of the ring of Casimir functions are given by
\begin{equation}
C_{k}~=~\mbox{tr}~ L^{k}~=~\mbox{tr}~\hat{L}^{k}.
\label{casimir}
\end{equation}

\section{Quantization}
\setcounter{equation}{0}

Let $A$ be a factorizable quasitriangular Hopf algebra \cite{dr,reshS}.
By definition, a quasitriangular  Hopf algebra is a Hopf algebra with a
distinguished element  $R\in A~\otimes~ A$ satisfying the following
properties:
\begin{enumerate}
\item[(i)] \begin{equation}
\Delta'(x)~=~R\Delta(x)R^{-1}
\label{deltaprime}
\end{equation}
 for any $x\in A$. (Here
$\Delta'$ denotes the opposite coproduct in $A$.)\\
\item[(ii)]
 \begin{equation}
(\Delta~\otimes~ id)R~=~R_{13}R_{23},~~~~~
(id~\otimes~\Delta )R~=~R_{13}R_{12},
\label{Rmatr}
\end{equation}
\item[(iii)] $$(\epsilon~\otimes~ id)R~=~(id~\otimes~\epsilon)R~=~1.$$
\end{enumerate}
Under these assumptions $R$ is invertible and $R^{-1}~=~(S~\otimes~
id)R~=~(id~\otimes~ S)R$, where $S$ is the antipode of $A$.

Let $\sigma$  be the permutation map in $A~\otimes~ A$. Put
$R_{+}~=~R,~R_{-}~=~\sigma(R^{-1})$. Let $A^{*}$  be the dual Hopf algebra
of $A$, and $A^{0}$  the same algebra with the  opposite coproduct. Let
$R^{\pm}:A^{0}\rightarrow A$  be the maps
\begin{equation}
R^{\pm}:~f~\mapsto~ <f~\otimes~ id~,~R_{\pm}>
\label{homo}
\end{equation}
Then by (\ref{deltaprime}) $R^{\pm}$ are Hopf algebra
homomorphisms. Let us
consider the combined mapping
\begin{equation}
A^{0}\stackrel{(R^{+}\otimes R^{-})\Delta^{0}}{\longrightarrow}A\otimes
A\stackrel{m(id\otimes S^{-1})}{\longrightarrow}A.
\label{qfact}
\end{equation}
A quasitriangular Hopf algebra is called {\em factorizable} if the
composition map is a linear space isomorphism (for infinite-dimensional
spaces we require that it is an isomorphism of an open dense subspace in
$A^{0}$ onto an open dense subspace in $A$).

Choose a linear basis $\{e_{i}\}$ in $A$ and let $\{e^{i}\}$ be the dual
basis in $A^{0}$. Let
\begin{equation}
T~=~e^{i}~\otimes~ e_{i}\in A^{0}~\otimes~ A
\label{T}
\end{equation}
be the canonical element. We shall write, using the standard tensor
notation,
\begin{eqnarray*}
T_{1}T_{2}~~=~~e^{i}e^{j}~\otimes~ e_{i}~\otimes~ e_{j}\in
A^{*}~\otimes~ A~\otimes~ A,\\
T_{2}T_{1}~~=~~e^{j}e^{i}~\otimes~ e_{i}~\otimes~ e_{j}\in
A^{*}~\otimes~ A~\otimes~ A .
\end{eqnarray*}
{}From
(\ref{deltaprime}) we have
\begin{equation}
T_{2}T_{1}~=~R~T_{1}T_{2}~R^{-1}.
\label{RTT}
\end{equation}
Put
\begin{equation}
L^{\pm}~=~(R^{\pm}~\otimes~ id)~T~\in~ A~\otimes~ A.
\label{L+-}
\end{equation}
We have
\begin{eqnarray}
L_{2}^{\pm}L_{1}^{\pm}~~=~~R_{+}~L_{1}^{\pm}L_{2}^{\pm}~R_{+}^{-1},
\nonumber\\
L_{2}^{-}L_{1}^{+}~~=~~R_{+}L_{1}^{+}L_{2}^{-}R_{+}^{-1}.
\label{RLL}
\end{eqnarray}
Put $L~~=~~L_{+}^{-1}~L_{-}$.Then we get from (\ref{RLL})
\begin{equation}
R_{+}^{-1}L_{2}R_{+}~L_{1}~~=~~L_{1}~R_{-}^{-1}L_{2}R_{-}.
\label{RLRL}
\end{equation}

Now let $\rho_{V},\rho_{W}$ be repesentations of $A$  in linear
spaces $V,W$. Put
\begin{eqnarray*}
T^{V}~~=~~(id~\otimes~\rho_{V})~T~\in~ A^{0}~\otimes~ \mbox{End}V,\\
L^{\pm V}~~=~~(id~\otimes~\rho_{V})~L_{\pm}~\in~ A~\otimes~\mbox{End}V.
\end{eqnarray*}
 Then relations (\ref{RTT}, \ref{RLL}, \ref{RLRL}) yield
 \begin{eqnarray}
 T_{2}^{W}T_{1}^{V}~~=~~R^{VW}T_{1}^{V}T_{2}^{W}R^{VW~-1},\nonumber\\
 L_{2}^{\pm ~W}L_{1}^{\pm ~V}~~=~~R^{VW}L_{1}^{\pm ~V}L_{2}^{\pm
 ~W}(R^{VW})^{-1},\nonumber\\
 L_{2}^{-~W}L_{1}^{+~V}~~=~~R^{VW}L_{1}^{+~V}L_{2}^{-~W}
 \mbox{$(R^{VW})$}^{-1},
\nonumber\\
\mbox{$(R_{+}^{VW})$}^{-1}L_{2}^{W}R_{+}^{VW}L_{1}^{V}~~=~~
 L_{1}^{V}\mbox{$(R_{-}^{VW})$}^{-1}L_{2}^{W}R_{-}^{VW}. \label{RTTrep}
 \end{eqnarray}

Assume now that $A$ is quasi-classical. This means that $A$  is a free
module over the ring $\bf{C}\mbox{$[[h]]$}$, where $h$ is a
deformation parameter,
and, moreover, $A/hA$  is isomorphic (as a Hopf algebra) to the
universal enveloping algebra of a Lie algebra $\frak{g}$. Formula
 (\ref{deltaprime}) then implies that
$$R_{\pm}~=~1+hr_{\pm}~+~o(h),$$
where $r_{\pm}\in U(\frak{g})^{\otimes 2}$. It is easy to see that in
fact $r_{\pm}\in\frak{g}~\otimes~\frak{g}$ where $\frak{g}$ is identified
with the subalgebra of primitive elements in $U(\frak{g})$. Formulae
 (\ref{Rmatr}) imply that $r_{\pm}$ satisfy the classical Yang---Baxter
identity. The factorization map (\ref{qfact}) induces an isomorphism of
linear spaces $\frak{g}^{*}\rightarrow \frak{g}$, i.e. an invariant inner
product on $\frak{g}$. Thus $\frak{g}$ has the structure of a factorizable
Lie bialgebra. Let us define the associated Poisson Lie groups $G,G^{*}$
and Poisson algebras $Fun(G),~ Fun(G^{*})$, as in Section 2. We shall
assume that $G,G^{*}$  are classical algebraic linear groups and that
an exact
representation $(\rho,V)$ of $G$ agrees with the representation
$\rho_{V}$ of $A$. The algebras $Fun(G), Fun(G^{*})$ are generated by
the matrix coefficients of matrices $\rho_{V}(T),
\rho_{V}(L^{\pm})$,respectively. There are obvious relations which
express the symmetry conditions for  $T\in G$  and describe the image of
$G^{*}$ in $G$ under the mappings $r_{\pm}$. (For instance, if $G~=~GL(n)$
given with its standard matrix representation, and r is the standard
r-matrix on $gl(n)$ associated with the Gauss decomposition , then the
symmetry conditions are void and $r_{\pm}(G)$ are the opposite Borel
subgroups in $GL(n)$.) All these relations may be explicitly quantized,
i.e. they admit canonical deformations which make them compatible with
the quantum commutation relations \cite{frt}. In the sequel, we  shall
not write down these relations explicitly. The reader may assume that
$\frak{g}~=~ gl(n)$, however, an extension to other classical Lie algebras
is always straightforward.

\begin{proposition}
Associative algebras $Fun_{q}(G), Fun_{q}(G^{*})$ generated by the
matrix coefficients of $T^{V}$, $L^{\pm~V}$ (or $L^{V}$) and relations
 (\ref{RTTrep}) (along, if necessary, with the quantum symmetry relations)
 are quantizations of the Poisson algebras $Fun(G),
Fun(G^{*})$,respectively.
\label{Prop3.1}
\end{proposition}
The generators $T^{V},L^{\pm~V}$ are expressed through the canonical
element
 (\ref{T}) and the universal R-matrix. This defines the
homomorphisms
\begin{equation}
Fun_{q}(G^{*})\rightarrow A,~~ Fun_{q}(G)\rightarrow
A^{*}.
\label{realize}
\end{equation}
For $A~=~U_{q}(G)$ we may use explicit formulae for the universal R-matrix
\cite{kr,ls} to express the generators $L^{\pm~V},T^{V}$ in terms of the
Drinfeld---Jimbo generators. (For $\frak{g}~=~sl_{2}$ these formulae are
elementary and are given in \cite{frt})

{\em Remark.}  One can check, using the results of \cite{frt} that
 the mappings
 (\ref{realize}) are actually isomorphisms (i.e. there are no extra
relations in the algebras)

The construction below makes sense for any Hopf algebra. Let us consider
an action $A~\otimes~ A^{*}\rightarrow A^{*}$ by 'left derivations`
\begin{equation}
x\otimes f\mapsto D_{x}f ~=~<x~\otimes~ id~,~ \Delta f>.
\label{der}
\end{equation}
Obviously, we have
\begin{equation}
<x~,~f>~=~\epsilon(D_{x}~f),
\label{counit}
\end{equation}
i.e the canonical pairing between $A$ and $A^{*}$ is given by the 'value
of the derivative at the unit element`. Let us consider an asssociative
algebra $\cal H$ of operators in $A^{*}$ generated by derivations
$D_{x},x\in A$,
 and by multiplication operators
\begin{equation}
m_{f}:\phi\mapsto f\phi, f\in A^{*}.
\label{mult}
\end{equation}

\begin{definition}
The algebra $\mbox{\cal H}$ is called the Heisenberg double of $A$.
\end{definition}
{\em Example.}If $A~=~S(V)$ is the symmetric algebra of  a linear
 space $V$,
$\cal H$ is the enveloping algebra of the Heisenberg algebra generated by
$V\oplus V^{*}$.

Assume again that A is factorizable.  We shall write down the
commutation relations for $\mbox{\cal H}(A)$ in the form that allows a
comparison with the Poisson algebra $Fun(D_{+})$. Let us introduce the
operator-valued matrices
\begin{eqnarray*}
D_{L^{\pm}}~~=~~(D~\otimes~ id)~~L^{\pm}~~\in~~\mbox{End}A^{*}~\otimes~ A,\\
m_{T}~~=~~(id~\otimes~ m)~~T~~\in~~ A~\otimes~\mbox{End}A^{*}.
\end{eqnarray*}
Obviously,
$$D:A\rightarrow \mbox{End}~A^{*}~,~m:A^{*}\rightarrow \mbox{End}~A^{*}$$
are homomorphisms of algebras, so it is sufficient to  compute the
commutation relations between the matrices $D_{L^{\pm}}$, or $D_{L}$
and $m_{T}$.

\begin{proposition}
We have
\begin{eqnarray}
D_{L_{1}^{\pm}}\circ T_{2}~=~T_{2}\circ D_{L_{1}^{\pm}}R_{\pm},\nonumber\\
 D_{L_{1}}\circ T_{2}~=~T_{2}\circ R_{+}D_{L_{1}}R_{-}^{-1},\nonumber\\
 <L_{1}^{\pm}, T_{2}>~=~R_{\pm}.
 \label{DLT}
 \end{eqnarray}
\label{Prop3.2}
 \end{proposition}
 In order to interprete these relations let us consider an associative
 algebra generated by the matrix coefficients of the matrices
 $T^{V},L^{\pm~V}$ (or $L^{V}$) and relations (\ref{RTTrep}) and the
 supplementary relations
   \begin{eqnarray}
   L_{1}^{\pm~V}T_{2}^{W}~~=~~T_{2}^{W}L_{1}^{\pm~V}R_{\pm}^{VW},
   \nonumber\\
   L_{1}^{V}T_{2}^{W}~~=~~T_{2}^{W}R_{+}^{VW}L_{1}^{V}(R_{-}^{VW})^{-1}.
  \label{LTR}
   \end{eqnarray}
  (We always tacitly assume that $L,T$ satisfy also the symmetry relations
  which hold for classical groups.)

  \begin{proposition} Assume that $A$ is a quasi-classical
  factorizable Hopf
  algebra, as in \ref{Prop3.1}.Then
  \begin{enumerate}
  \item[(i)] The algebra (\ref{RTTrep}, \ref{LTR}) is a
  quantization of the Poisson
  algebra $Fun(D_{+})$.\\
  \item[(ii)] Realization of $L,T$ by derivation and multiplication
  operators
  in $A^{*}$ gives an exact representation of this algebra in
  $\mbox{End}A^{*}$.
  \end{enumerate}
  \label{Prop3.3}
  \end{proposition}
Alternatively,  we can use another set of generators of the Heisenberg
double for which the commutation relations have a more symmetric form.

\begin{proposition}
Put $X~=~L_{+}T^{-1},~~Y~=~L_{-}T^{-1}$. Then the following relations
hold:
\begin{eqnarray}
R_{+}X_{1}X_{2}~=~X_{2}X_{1}R_{-}^{-1},\nonumber\\
R_{+}Y_{1}Y_{2}~=~Y_{2}Y_{1}R_{-}^{-1},\nonumber\\
R_{+}X_{1}Y_{2}~=~Y_{2}X_{1}R_{+}^{-1}.
\label{RXY}
\end{eqnarray}
The algebra generated by the matrix coefficients of $X,Y$ and relations
(\ref{RXY}) is a quantization of the Poisson algebra (\ref{xybr}).
\label{Prop.3.4}
\end{proposition}

Along with the Heisenberg double one can also define the {\em Drinfeld
double}. Recall that the  double of a Lie bialgebra is a factorizable
Lie bialgebra, and the associated classical r-matrices are essentially
the projection operators. In complete analogy, the Drinfeld double $\cal
D$ $(A)$ of a Hopf algebra $A$  is a canonically defined factorizable Hopf
 algebra with the underlying linear space $A~\otimes~ A^{*}$.
 \newtheorem{Thm}{Theorem (Drinfeld, \cite{dr})}

 \begin{Thm}
 There exists a  unique structure of a
 \hyphenation{qua-si-tri-an-gu-lar} Hopf algebra on
 $\cal{D}$ $(A) ~=~A~\otimes~ A^{*}$ such that
 \begin{enumerate}
 \item[(i)] $A,A^{*}\subset \cal D $ are subalgebras in $\cal D$.\\
 \item[(ii)] As a coalgebra, $\cal D$ $ ~=~ A~\otimes~ A^{0}$.\\
\item[(iii)] The universal R-matrix is the image of the canonical
element $T\in A~\otimes~
 A^{*}$  (cf. (\ref{T})) under the natural embedding $A\otimes A^{*}
 \hookrightarrow \cal D ~\otimes~\cal D$.
\end{enumerate}
The Hopf algebra $\cal D$ is factorizable (\cite{reshS}).
 \label{THM}
 \end{Thm}
It is interesting to observe that $\cal D$ $(A)$ also admits an operator
realization. Let us first define the (right) adjoint action of an
arbitrary Hopf algebra on itself:
\begin{equation}
\mbox{Ad}:A~\otimes~ A\rightarrow A:x~\otimes~ y\mapsto
Sy^{(1)}xy^{(2)},
\label{ad}
\end{equation}
where $\Delta y~=~y^{(1)}~\otimes~ y^{(2)}$ is the coproduct in $A$. The
{\em coadjoint action} $\mbox{Ad}^{*}:A^{*}~\otimes~ A\rightarrow A^{*}$
is defined by
\begin{equation}
<\mbox{Ad}^{*}(f~\otimes~ x)~,~y>~=~<f~,~\mbox{Ad}(x~\otimes~
y)>.
\label{defcoad}
\end{equation}
{}From the definition one easily gets that
\begin{equation}
\mbox{Ad}^{*}(f~\otimes~ x)~=~<Sf^{(1)}f^{(3)},x>f^{(2)},
\label{coad}
\end{equation}
where $\Delta ^{(2)} f~=~f^{(1)}~\otimes~ f^{(2)}~\otimes~ f^{(3)}$.
Let us consider the algebra $\cal D$ of operators acting on
$A^{*}$ generated by
multiplication operators $m_{f},f\in A^{*}$ and the operators
$\mbox{Ad}^{*}x,x\in A$.

\begin{proposition}
As an associative algebra, $\cal D$ is isomorphic to the Drinfeld double
of $A$.
\label{Prop3.5}
\end{proposition}

Let us assume now that $A$ is a quasiclassical factorizable  Hopf
algebra. We shall describe its Drinfeld double as a quantization of a
Poisson algebra. The quantum duality principle suggests that $\cal D$ $(A)
\simeq Fun_{q}(D^{*})$ where $D^{*}$ is the dual double of
$G,G^{*}$. We shall see now that this is indeed the case. Let
$T\in~A^{*}~\otimes~ A$ be the canonical element (see (\ref{T})); put
$L^{\pm}~=~(R^{\pm}~\otimes~ id)T$. We fix a representation $(\rho,V)$ and
set
\begin{eqnarray*}
\hat{\cal L}^{\pm~V}~=~(Ad^{*}~\otimes~ \rho_{V})L^{\pm},\\
\hat{\cal T}^{V}~=~(m ~\otimes~ \rho_{V})T.
\end{eqnarray*}
(Here $m:A^{*}\rightarrow \mbox{End}A^{*}$ is the representation of
$A^{*}$ by multiplication operators, as in (\ref{mult}).)

\begin{proposition}
\begin{enumerate}
\item[(i)] The operator matrices $\hat{\cal L}^{\pm~V},\hat{\cal L}^{V}$
satisfy
\begin{equation}
\hat{\cal L}_{2}^{\pm~V}\hat{\cal T}^{V}_{1}~=~R^{VV}_{\pm}\hat{\cal
T}^{V}_{1}\hat{\cal L}^{\pm~V}_{2}(R_{\pm}^{VV})^{-1}.
\label{3.21}
\end{equation}
\item[(ii)] The associative algebra generated by the matrix
coefficients of
$\hat{\cal L}^{\pm~V}$ ,$\hat{\cal T}^{V}$  and relations (\ref{3.21},
 \ref{RTTrep})
is a quantization of the Poisson algebra $Fun(D^{*})$ with the Poisson
bracket relations (\ref{double*})
\item[(iii)] The coproduct in $\cal D$ $(A)$ is given by
\begin{eqnarray*}
\Delta\hat{\cal T}~=~\hat{\cal T}_{2}~\dot{\otimes}~\hat{\cal T}_{1},
\\
\Delta\hat{\cal L}^{\pm}~=~\hat{\cal L}^{\pm}_{1}\dot{~\otimes~}\hat{\cal
L}^{\pm}_{2}.\end{eqnarray*}
\end{enumerate}
\label{Prop3.8}
\end{proposition}

In a similar way, the algebra $Fun_{q}(D_{-})$ may be identified with
the {\em dual}  of the Drinfeld double.

\begin{proposition}
\begin{enumerate}
\item[(i)] The associative algebra $Fun_{q}(D_{-})$ generated by matrix
coefficients of the matrices $\cal L^{\pm},~\cal T$ , relations
 (\ref{RTT},\ref{RLL}) and the supplementary relation
\begin{equation}
\mbox{$\cal T$}_{1}\mbox{$\cal L$}^{\pm}_{2}~~=
{}~~\mbox{$\cal L$}^{\pm}_{2}\mbox{$\cal T$}_{1}
\label{3.15}
\end{equation}
is isomorphic to the dual of the Drinfeld double $\cal D$ $(A)^{*}$.
\item[(ii)] the algebra $Fun_{q}(D_{-})$ is a quantization of the Poisson
algebra $Fun(D_{-})$ with relations (\ref{doublebr}).
\end{enumerate}
\label{Prop3.9}
\end{proposition}

In order to describe the coproduct in $Fun_{q}(D_{-})$  it is convenient
to introduce another system of generators suggested by (\ref{xybr}). Put
\begin{equation}
\check{X}~~=~~\mbox{$\cal{L}^{+}\cal{T}$},~~~~~\check{Y}~~=
{}~~\mbox{$\cal{L}^{-}\cal{T}$}.
\label{3.16}
\end{equation}

\begin{proposition}
\begin{enumerate}
\item[(i)] The matrix coefficients of $\check{X},\check{Y}$ satisfy the
following commutation relations
\begin{eqnarray}
\check{X}_{2}\check{X}_{1}~~=~~R_{+}\check{X}_{1}\check{X}_{2}R_{+}^{-1},
\nonumber\\
\check{Y}_{2}\check{Y}_{1}~~=~~R_{+}\check{Y}_{1}\check{Y}_{2}R_{+}^{-1},
\nonumber\\
\check{Y}_{2}\check{X}_{1}~~=~~R_{+}\check{X}_{1}\check{Y}_{2}R_{+}^{-1}.
\label{3.17}
\end{eqnarray}
\item[(ii)] Formulae
$$\Delta\check{X}~~=~~\check{X}\dot{~\otimes~}\check{X},~~~~
\Delta\check{Y}~~=~~\check{Y}\dot{~\otimes~}\check{Y}$$
 define on
$Fun_{q}(D_{-})$ the coalgebra structure which is in agreement with that
of $\mbox{\cal D}(A)^{*}$.
\item[(iii)] Let $X,Y$ be the matrices generating $\mbox{\cal H}(A) $
and satisfying
relations (\ref{RXY}). Formulae
$$C(X)~~=~~\check{X}\dot{~\otimes~}\check{X},~~C(Y)~~=
{}~~\check{Y}~\dot{\otimes}~\check{Y}$$
define on $\mbox{$\cal H$}(A)$ the structure of a left
$Fun_{q}(D_{-})$-comodule.
\end{enumerate}
\label{Prop3.10}
\end{proposition}

Let us now return to the study of the Heisenberg double. We shall
describe the quantum analogue of the Poisson reduction and the dual pair
described in (\ref{diagr}). We have just mentioned that
$\mbox{$\cal H$}(A)$ has a
natural structure of a left $\mbox{$\cal D$}(A)^{*}$-comodule.
In a similar way,
one can define on $\mbox{$\cal H$}(A)$ the structure of a right
$\mbox{$\cal
D$}(A)^{*}$-comodule. We may specialize these formulae to get on
$\mbox{$\cal
H$}(A)$  the structure of a left and right $Fun_{q}(G)$-comodule. Namely,
put
\begin{eqnarray}
C_{L}(X)~~=~~T~\dot{\otimes}~X,~~C_{L}(Y)~~=~~T~\dot{\otimes}~Y,
\nonumber\\
C_{R}(X)~~=~~X~\dot{\otimes}~T^{-1},~~C_{R}(Y)~~=~~Y~\dot{\otimes}~T^{-1},
\label{3.22}
\end{eqnarray}
where $T$ is the generator matrix for $Fun_{q}(G),~T^{-1}~~=~~(S\otimes
id)T$, and $S$ is the antipode of $Fun_{q}(G)$. The necessary formal
properties of $C_{L},C_{R}$  are easily verified.

Let now $Fun_{q}(G\backslash D),~ Fun_{q}(D/G)\subset Fun_{q}(D_{+})$ be
the subalgebras of left (right) coinvariants. By definition,
\begin{eqnarray*}
\phi\in Fun_{q}(g\backslash D)\Longleftrightarrow C_{L}((\phi)\in
1~\otimes~ Fun_{q}(D_{+}),\\
{}~~\psi\in Fun_{q}(D/G)\Longleftrightarrow C_{R}(\psi)\in
Fun_{q}(D_{+})~\otimes~ 1.
\end{eqnarray*}

\begin{proposition}
\begin{enumerate}
\item[(i)] The algebra $Fun_{q}(G\backslash D)$ is generated by the matrix
coefficients of $L'~~=~~X^{-1}Y$; in a similar way, the algebra
$Fun_{q}(D/G)$ is generated by $L~~=~~XY^{-1}$.
\item[(ii)] The matrices $L,L'$  satisfy the commutation relations
(\ref{RLRL});
thus $$Fun_{q}(D/G)\simeq Fun_{q}(G\backslash D)\simeq Fun_{q}(G^{*})$$.
\item[(iii)] Moreover, $L_{1}L'_{2}~~=~~L'_{2}L_{1}$, i.e. the subalgebras
$Fun_{q}(D/G),~Fun_{q}(G\backslash D)$ centralize each other in
$Fun_{q}(D_{+})\simeq \mbox{$\cal H$}(A)$.
\end{enumerate}

\label{Prop3.11}
\end{proposition}

Clearly, the algebra $Fun_{q}(D/G)$ inherits the structure of a left
$Fun_{q}(G)$-comodule. It is given by
\begin{equation}
C(L)~~=~~T_{1}L_{2}T_{1}^{-1}.
\label{3.23}
\end{equation}
Formula (\ref{3.23}) gives in the present case a quantum analogue of
'dressing transformations`. The next assertion is similar to the
description of the center of the Poisson algebra $Fun(D/G)$; in the same
time, it provides a quantum version of the well known  Gelfand theorem
describing the center of a  universal enveloping algebra.

\begin{proposition}
The center of $Fun_{q}(D/G)$  coincides with the subalgebra of
coinvariants of the coaction (\ref{3.23}).
\label{Prop3.12}
\end{proposition}

We shall study quantum dressing transformations more thoroughly in a
separate paper.

\section{The Heisenberg Double and the Quantum Fourier Transform}
\setcounter{equation}{0}

So far we have studied the commutation relations for $L^{\pm~V},T^{V}$
fixing the representation $(\rho, V)$ of our Hopf algebra. If, instead,
we consider arbitrary irreducible representations of $A$ and $A^{*}$ and
their tensor products, we may relate the quantum duality principle with
the intuitively attractive quantum Fourier transform.\footnote{A
different version of the quantum Fourier transform is dicussed in
\cite{LubMad}.}

Let $A$ be a Hopf algebra, $A^{*}$ its dual. Let
$\hat{A}~=~\mbox{Spec}(A)$  be a set of its irreducible representations.
We do not fix a $*$-structure on $A$ and hence do not assume that
rerpresentations $\pi \in\hat{A}$ are unitary. For $\pi\in\hat{A}$ let
$V_{\pi}$ be the corresponding $A$-module. Let $Fun(\hat{A})$ be the
space of functions on $\hat{A}$  such that
$\phi_{\lambda}\in\mbox{End}V_{\lambda}$; $Fun(\hat{A})$ is an algebra
with respect to pointwise multiplication. It is natural to identify the
dual space $Fun(\hat{A})^{*}$ with the space of {\em matrix-valued
measures} on $\hat{A}$. For
 $\phi\in Fun(\hat{A}),~ \sigma~=~d\sigma(\lambda)\in
Mes(\hat{A})$ we put \footnote{In our exposition we shall ignore all
convergence questions, so the reader may assume that all representations
are finite-dimensional.}
$$<\sigma,\phi>~=~\int_{\hat{A}}\mbox{tr}_{V_{\lambda}}
(\phi(\lambda)d\sigma(\lambda)).$$

We shall assume that the set  $\hat{A}$ is closed with respect to the
tensor product. More precisely, we suppose that for any
$\lambda_{1},\lambda_{2}\in\hat{A}$
$$V_{\lambda_{1}}~\otimes~ V_{\lambda_{2}}\simeq
\int\bigoplus_{\lambda\in\hat{A}}V_{\lambda}~\otimes~ W_{\lambda},$$
where $W_{\lambda}$ is the multiplicity space.

Let us introduce the Clebsch---Gordan coefficients by the formula
$$\lambda_{1}~\otimes~\lambda_{2}(x)~=~
\int_{\hat{A}}\mbox{tr}_{V_{\lambda}}(\lambda(x)
dC(\lambda\mid\lambda_{1},\lambda_{2})).$$
By definition, $dC(\lambda\mid\lambda_{1},\lambda_{2})$ is a function of
variables $\lambda_{1},\lambda_{2}$ and a measure in variable
$\lambda$ with values in $\mbox{End}(V_{\lambda_{1}}~\otimes~
V_{\lambda_{2}}~\otimes~ V_{\lambda})$ which commutes with all
intertwiners
$I\in \mbox{End}_{A}(V_{\lambda_{1}}~\otimes~
V_{\lambda_{2}}~\otimes~ V_{\lambda})$.

In a similar way we define the set $\hat{A}^{*}~=~\mbox{Spec}(A^{*})$ and
the spaces $Fun(\hat{A}^{*}),~ Mes(\hat{A}^{*})$. Let us denote by
$\cal R$
the canonical element in $A^{*}~\otimes~ A$ \footnote{We are changing
slightly the notation introduced in Section 3. Indeed, $\cal R$ coincides
with the universal R-matrix for $\mbox{\cal D}(A)$.}  and put
\begin{eqnarray}
T_{\rho}~~=~~(id~\otimes~\rho)\mbox{$\cal R$},~~\rho\in\hat{A},\nonumber\\
L_{\lambda}~~=~~(\lambda~\otimes~ id)\mbox{$\cal R$},
{}~~\lambda\in\hat{A}^{*}.
\label{4.3}
\end{eqnarray}
Define the {\em spectral representation} $T:~Mes(\hat{A})\rightarrow A$ by
\begin{equation}
T(\phi)~~=~~\int_{\hat{A}}\mbox{tr}_{V_{\rho}}(T_{\rho}d\phi(\rho)).
\label{4.4}
\end{equation}
In a similar way, the spectral representation
$L:~Mes(\hat{A}^{*})\rightarrow A$ is defined by
\begin{equation}
L(\psi)~=~\int_{\hat{A}^{*}}\mbox{tr}(L_{\lambda}d\psi(\lambda)).
\label{4.5}
\end{equation}
We define the convolutions of measures
\begin{eqnarray}
*:~Mes(\hat{A})~\otimes~ Mes(\hat{A})\rightarrow Mes(\hat{A}),\nonumber\\
*:~Mes(\hat{A}^{*})~\otimes~ Mes(\hat{A}^{*})\rightarrow
Mes(\hat{A}^{*}),
\label{conv}
\end{eqnarray}
demanding that
\begin{eqnarray}
T(\phi_{1}*\phi_{2})~=~T(\phi_{1})T(\phi_{2}),\nonumber\\
L(\psi_{1}*\psi_{2})~=~L(\psi_{1})L(\psi_{2}).
\label{4.6}
\end{eqnarray}

\begin{proposition}
\begin{enumerate}
 \item[(i)] The convolution of measures on $\hat{A}$ is given by
 $$\phi_{1}*\phi_{2}(\rho)~=~\int_{\hat{A}\times\hat{A}}\mbox{tr}_
 {V_{\rho_{1}}~\otimes~
 V_{\rho_{2}}}C(\rho\mid\rho_{1},\rho_{2})
 d\phi_{1}(\rho_{1})d\phi_{2}(\rho_{2}),$$
 where $C(\rho\mid\rho_{1},\rho_{2})$ are the Clebsch---Gordan
 coefficients of $A$.
 \item[(ii)] In a similar way, the convolution of measures on
 $\hat{A}^{*}$ is
 given by
 $$\psi_{1}*\psi_{2}(\lambda)~=~ \int_{\hat{A^{*}}\times\hat{A}^{*}}
 \mbox{tr}_{W_{\lambda_{1}}~\otimes~
 W_{\lambda_{2}}}C^{*}(\lambda\mid\lambda_{1},\lambda_{2})
 d\psi_{1}(\lambda_{1})d\psi_{2}(\lambda_{2}),$$
 where $C^{*}(\lambda\mid\lambda_{1},\lambda_{2})$  are the
 Clebsch---Gordan coefficients of $A^{*}$.
 \end{enumerate}
\label{Prop4.1}
\end{proposition}

  In terms of the spectral representations the coupling
   $A^{*}~\otimes~ A\rightarrow \bf{C}$ takes the form
   \begin{equation}
   <\phi\mid\psi>~=~\int\mbox{tr}_{V_{\rho}~\otimes~ W_{\lambda}}
   (d\phi(\rho)R_{\rho\lambda}d\psi(\lambda)),
   \label{4.7}
   \end{equation}
   where
   \begin{equation}
   R_{\rho\lambda}~~=~~(\rho~\otimes~\lambda)\cal R.
   \label{4.8}
   \end{equation}

Let us define the Fourier transform
$$\phi:~~Mes(\hat{A}^{*})\rightarrow Fun(\hat{A})$$
by
\begin{equation}
\Phi\phi(\lambda)~~=~
{}~\int\mbox{tr}_{V_{\lambda}}(d\phi(\rho)R_{\rho\lambda}).
\label{4.9}
\end{equation}

\begin{proposition}
We have
 $$\Phi(\phi_{1}*\phi_{2})~=~\Phi\phi_{1}\Phi\phi_{2}.$$
\label{Prop4.2}
\end{proposition}
In a similar way we define the conjugate Fourier transform
$$\hat{\Phi}:~Mes(\hat{A})\rightarrow Fun(\hat{A}^{*}).$$

{\em Example.}  Let $A~=~S(V)$ be the symmetric algebra of a linear space
$V$. Then $A^{*}\simeq S(V^{*})$. Choose a basis $e_{i}$ in $V$  and let
$e^{i}$ be the dual basis in $V^{*}$ . Then the canonical element in
$A^{*}~\otimes~ A$ is given by
$$\cal R~=~\rm{exp}(~\sum e^{i}~\otimes~ e_{i}).$$
Irreducible representations of $A$ and $A^{*}$ are 1-dimensional and we
have $\hat{A}~=~V^{*}$, $\hat{A}^{*}~=~V$. The convolutions
(\ref{4.6}) become
ordinary convolutions of measures on a linear space, and the Fourier
transform is the ordinary Fourier---Laplace transform on $V$.

Let us define an action
$$Mes(\hat{A}^{*})~\otimes~ Mes(\hat{A})\rightarrow Mes(\hat{A})$$
by the formula
\begin{equation}
\phi~\otimes~\psi\rightarrow \Phi\phi\cdot\psi.
\label{MULT}
\end{equation}

\begin{proposition}
The algebra of operators acting on $Mes(\hat{A})$ which is generated by
convolution operators (\ref{conv}) and multiplication operators
(\ref{MULT})
is isomorphic to the Heisenberg double $\mbox{$\cal H$}(A)$.
\label{Prop4.3}
\end{proposition}

\section{The Twisted Double, and Deformations of Classical and
Quantum Algebras}

Let us return to the study of Poisson algebras of functions and consider
certain types of their deformations. We begin with a simple example:
deformations of the Lie---Poisson bracket on the dual space of a Lie
algebra.Let $\frak{g}$ be a Lie algebra such that $H^{2}(\frak{g})\neq 0$.
Fix a nontrivial cocycle $\omega\in C^{2}(\frak{g})$, and let $\hat{\frak
g}_{\omega}~=~\frak{g}\oplus\bf{R}$ be the associated central extension of
$\frak{g}$. The dual space $\hat{\frak{g}}_{\omega}^{*}$ is naturally
isomorphic to $\frak{g}^{*}\oplus\bf{R}$. Let $e$ be an affine coordinate
on  $\bf{R}$.  Clearly, $e$  lies in the center of the Poisson algebra
$Fun(\hat{\frak{g}}_{\omega}^{*})\simeq Fun(\frak{g})~\otimes~
Fun(\bf{R})$.
Fixing the value of $e$ we get a 1-parameter family of Poisson brackets
on $\frak{g}$ which may be regarded as a deformation of the original
Lie---Poisson bracket. Its 'universal` deformation is parametrized by
the elements of $H^{2}(\frak{g})$.

Let us consider similar deformations for general Poisson Lie groups. We
shall again begin with the deformations associated with central
extensions (cf. \cite{reshS2}).

Let $(\frak{g}, \frak{g}^{*})$ be a factorizable Lie bialgebra,
$\partial\in\mbox{Der}(\frak{g},\frak{g}^{*})$ its derivation. By
definition, this means that $\partial$ is a derivation of $\frak{g}$
which is skew with respect to the inner product on $\frak{g}$  and
commutes with $r\in\mbox{End}\frak{g}$ . Formula
\begin{equation}
\omega(X,Y)~=~(X~,~\partial Y)
\label{5.1}
\end{equation}
defines a 2-cocycle on $\frak{g}$ . Thus we get an embedding
$\mbox{Der}(\frak{g},\frak{g}^{*})\hookrightarrow C^{2}(\frak{g})$.

{\em Example.} In typical applications $\frak{g}=L\frak a$ is a
 loop algebra equipped with
a standard r-matrix, $\partial ~=~\partial_{x}$ is the derivative in loop
parameter. In this case  it is easy to see that the class group
$[\rm{Der}(\frak{g},\frak{g}^{*})]$ (i.e. the quotient group of all
derivations modulo inner derivations) is isomorphic to $H^{2}(L\frak
a)$. For a simple Lie algebra $\frak a$ the group $H^{2}(L\frak a)\simeq
\bf{R}$ is generated by the cocycle (\ref{5.1}).

Let $\hat{\frak{g} }~=~\frak{g}\oplus \bf{R}$ be the central extension of
$\frak{g}$ associated with the cocycle (\ref{5.1}). Then $\hat{\frak{g}}$
has  a canonical structure of a Lie bialgebra; the dual Lie algebra
$\hat{\frak{g}^{*}}$ is the semi-direct product
$$\hat{\frak{g}}^{*}~=~\frak{g}^{*}\uplus \bf{R}\partial.$$
More precisely, let us define the commutator in $\hat{\frak{g}}^{*}$ by
\begin{equation}
[f+\alpha\partial~,~g+\beta\partial]~=~[f~,~g]_{*}~+~\alpha\partial
r(g)~-~\beta\partial r(f).
\label{5.2}
\end{equation}

\begin{proposition}
The pair $(\hat{\frak{g}},\hat{\frak{g}}^{*})$ is a  Lie bialgebra.
\label{Prop5.1}
\end{proposition}
It is easy to describe the double of $(\hat{\frak{g}},\hat{\frak
g}^{*})$. We shall do it under the simplifying assumption
\begin{equation}
\partial~-~\partial\cdot r^{2}~=~0.
\label{5.3}
\end{equation}
(This condition holds for standard r-matrices on loop algebras.)

Put
   $$\frak{d} ~~=~~\frak{g}\oplus\frak{g},
   ~~~~ \hat{\frak{d}}~~=~~\frak{d}\oplus \bf{R}\cdot
   c\oplus\bf{R}\cdot\partial.$$
   We define the inner product on $\hat{\frak{d}}$ by
   $$\begin{array}{ll}
\ll(X_{1},Y_{1},\alpha_{1},\beta_{1})~,
   ~(X_{2},Y_{2},\alpha_{2},\beta_{2})\gg~~=\\
{}~~<X_{1},X_{2}>-<Y_{1},Y_{2}>+
   \alpha_{1}\beta_{2}+\alpha_{2}\beta_{1}.
\end{array}$$
   Extend the derivation $\partial$ to $\frak{d}$  by
   $\hat{\partial}(X,Y)~~=~~(\partial X,-\partial Y)$
   and define the cocycle
   $\omega_{\frak{d}}$ which gives the 'central ` component of the Lie
   bracket on $\hat{\frak{d}}$ by
\begin{equation}
\omega_{\frak{d}}(a,b)~=~\ll a,\hat{\partial}b\gg.
\label{5.4}
\end{equation}

\begin{proposition}
Assume that condition (\ref{5.3}) holds. Then $\hat{\frak{d}}$ is
isomorphic to the double of $(\hat{\frak{g}},\hat{\frak{g}}^{*})$;
embeddings $\frak{g},\frak{g}^{*}\hookrightarrow \hat{\frak{d}}$ are given
by
$$(X,~\alpha)~\mapsto~(X,~X,~\alpha c)~~,
{}~~(f,~\beta)~\mapsto ~(r_{+}f,~r_{-}f
,~\beta~\partial).$$
\label{Prop5.2}
\end{proposition}
Let $\Gamma$ be the group of automorphisms of $(\frak{g},\frak{g}^{*})$
generated by $\partial$. The Lie group which corresponds to $\hat{\frak
g^{*}}$ is the semi-direct product $G^{*}\bowtie\Gamma$. According to
the quantum duality principle, the quantized universal enveloping
algebra $U_{q}(\hat{\frak{g}})$ may be identified with
$Fun_{q}(\hat{G}^{*})$. As a first step towards the study of this
algebra let us consider the Poisson algebra $Fun(\hat{G}^{*})$.

\begin{proposition}
Let $e$ be an affine coordinate on $\Gamma\simeq\bf{R}^{\times}$. Then
$e$ lies in the center of the Poisson algebra $Fun(\hat{G}^{*})$
\label{Prop5.3}
\end{proposition}
The function $e$ has the meaning of {\em central charge}. Proposition
(\ref{5.3}) means that the bracket on $\hat{G}^{*}$ depends on $e$ as a
parameter, i.e. we get a 1-parameter family $\{,\}_{e}$ of Posson
brackets on $G^{*}$.  After quantization we get a  1-parameter family of
algebras $Fun_{q}(G^{*})_{e}\simeq U_{q}(\hat{\frak{g}})/(e=\mbox{const})$
which may be regarded as quotient algebras of $U_{q}(\hat{\frak{g}})$
obtained by 'setting $e=\mbox{const}$`.

The algebra $Fun(G^{*})$ may be obtained from a larger algebra
$Fun(D_{+})$
by reduction. In a similar way, the algebras $Fun(G^{*})$ may be
obtained by reduction from the algebra of functions on the {\em twisted
double} of $(G,G^{*})$.

Remarkably, the definition of the twisted double and the Poisson
structure on $G^{*}_{-}~=~G^{*}\times\{e\}\subset\hat{G}^{*}$ depends not
on the derivation $\partial$ itself, but only on the automorphism
$\mbox{exp}(e\partial)\in\Gamma$. This allows to define the twisted
double and the twisted bracket on $G^{*}$ in a more general setting.

Put $D~=~G\times G$ . Let $\tau\in \mbox{Aut}(G)$  be an automorphism of
$G$. We assume that the corresponding automorphism of $\frak{g}$
preserves the inner product in $\frak{g}$ and commutes with $r$ .Put
\begin{eqnarray}
\hat{\tau}~=~\left(
\begin{array}{cc}
1&0\\
0&\tau
\end{array}
\right)\in\mbox{Aut}(G\times G),\nonumber\\
^{\tau}r_{\frak{d}}~=~\hat{\tau} r_{\frak
d}\hat{\tau}^{-1}~=~\left(
\begin{array}{cc}
r& -2r_{+}\circ\tau^{-1}\\
2\tau\circ r_{-}& -r
\end{array}
\right),
\label{5.6}
\end{eqnarray}
and define the twisted Poisson bracket on $D$ by
\begin{equation}
2\{\phi,\psi\}_{\tau}~=~\ll r_{\frak{d}}X,Y\gg~+~\ll^{\tau}r_{\frak
d}X',Y'\gg .
\label{5.7}
\end{equation}
(Here $X,Y,X',Y'$ are the left and right gradients of $\phi,\psi$,
respecxtively, defined by (\ref{grad}).) Note that the Jacobi identity for
(\ref{5.7}) follows from the classical Yang---Baxter identity
(\ref{cybe}) for
$r_{\frak{d}}$.

The group $D$ equipped with the Poisson bracket (\ref{5.7}) is called the
twisted double and is denoted by $D_{\tau}$. Assume, as in Section 2,
that $G$ is a classical matrix group with exact representation
$(\rho,V)$. The algebra  $Fun(D_{\tau})$ is generated by the matrix
coefficients of $\rho(x),\rho(y)$. The Poisson bracket relations for the
generators are given by
\begin{eqnarray}
\{x_{1},x_{2}\}_{\tau}~=
{}~\frac{1}{2}~(r^{VV}x_{1}x_{2}~+~x_{1}x_{2}r^{VV}),
\nonumber\\
\{y_{1},y_{2}\}_{\tau}~=
{}~\frac{1}{2}~(r^{VV}y_{1}y_{2}~+~y_{1}y_{2}r^{VV}),
\nonumber\\
\{y_{1},x_{2}\}_{\tau}~=
{}~-r_{+}^{VV}y_{1}x_{2}~-~y_{1}x_{2}^{\tau}r_{+}^{VV},
\label{5.8}
\end{eqnarray}
 where
\begin{eqnarray*}
^{\tau}r_{+}~=~(id~\otimes~\tau)r_{+}\in\frak{g}~\otimes~\frak g,\\
r^{VV}~=~\rho~\otimes~\rho (r)\in\mbox{End}(V~\otimes~ V).
\end{eqnarray*}

Let $G^{*}$ be the dual group of $G$ . As in Section 2, we identify it
with a subgroup in $D$.
\begin{proposition}
\begin{enumerate}
 \item[(i)] The actions of $G$ on $D_{\tau}$ by left and right
 translations
 defined by
 \begin{eqnarray*}
G\times D\rightarrow D:~~~g(x,y)~=~ (gx,gy),\\
 D\times G\rightarrow D:~~~(x,y)g~=~ (xg,yg^{\tau})
\end{eqnarray*}
 are admissible and form a dual pair.
 \item[(ii)] In a similar way, the actions of $G^{*}$ on $D_{\tau}$
 defined by
\begin{eqnarray*}
G^{*}\times D\rightarrow D:~~~h(x,y)~=~ (h_{+},h_{-}y),\\
 D\times G^{*}\rightarrow D:~~(x,y)h~=~(xh_{+},yh^{\tau}_{-})
\end{eqnarray*}
 are also admissible and form a dual pair.
 \end{enumerate}
\label{Prop5.4}
 \end{proposition}

The quotient spaces $D/G,G\backslash D$ may be identified with $G$ ,
the projection maps $p,p'$ being given by
$$p:D\rightarrow G:(x,y)\mapsto y^{-1}x,~~~~p':D\rightarrow G:
(x,y)\mapsto x(\mbox{$y^{-1}$})^{\bar{\tau}}.$$
(We write $\bar{\tau}=\tau^{-1}$ for shortness.) The reduced Poisson
bracket on $G$ is given by
\begin{equation}
\{L_{1},L_{2}\}_{\tau}~=~L_{1}r^{\tau}_{+}L_{2}~+~L_{2}r^{\tau}_{-}L_{1}
{}~-~\frac{1}{2}~(r~L_{1}L_{2}+L_{1}L_{2}~r),
\label{5.9}
\end{equation}
where $r^{\tau}_{+}=(id~\otimes~\tau)r_{+},~r^{\tau}_{-}=(\tau~\otimes~
id)r_{-}$ (note that $(\tau~\otimes~\tau)r~=~r$, due to the condition
imposed on $\tau$).

Another description of the reduced spaces $D/G,~D/G^{*}$ results from a
factorization problem on $D$. Observe that almost all elements $(x,y)\in
D\times D$ may be represented in the form
\begin{equation}
\begin{array}{ll}
(x,y)~=~(L_{+},L_{-})(T,T^{\tau})^{-1},\\
{}~(L_{+},L_{-})\in G^{*}\subset D,
{}~T\in G.
\end{array}
\label{5.10}
\end{equation}
Thus we may identify (up to a set of positive codimension)
$D/G^{*}\simeq G,~~G\backslash D\simeq G^{*}$ and compute the reduced
brackets for the generators of the algebras $Fun(G),~ Fun(G^{*})$. The
reduced bracket on $D/G^{*}\simeq G$  is the usual {\em Sklyanin
bracket}
$$\{T_{1},T_{2}\}~~=~~\frac{1}{2}[r~,~T_{1}T_{2}]$$
and does not depend on twisting. The brackets on $G^{*}\simeq
G\backslash D$ are given by
\begin{eqnarray}
\{L_{1}^{\pm}~,~L_{2}^{\pm}\}_{\tau}~~=~~
\frac{1}{2}~[r~,~L_{1}^{\pm}L_{2}^{\pm}],\nonumber\\
{}~\{L_{1}^{+}~,~L_{2}^{-}\}_{\tau}~~=~~r_{+}L_{1}^{+}L_{2}^{-}~-
{}~L_{1}^{+}L_{2}^{-}r_{+}^{\tau}.\label{5.11}
\end{eqnarray}
The identification of the two models for the quotient space
$G\backslash D$ is given by the {\em twisted factorization}
$$L~~=~~(L^{+})^{-1}(L^{-})^{\tau}.$$
Finally, using the expansion (\ref{5.10}) we may express the bracket on
$D_{\tau}$ in terms of the generators $L^{\pm},T$. It is sufficient to
compute only the brackets $\{L_{1}^{\pm},T_{2}\}$. The computation gives
$$\{L_{1}^{\pm}~,~T_{2}\}~~=~~L_{1}^{\pm}T_{2}r_{\pm},$$
i.e. the same formula as in the non-twisted case.

Assume now that $\tau~=~\exp(e\partial)$ where $\partial$  is a derivation
of $(\frak{g},\frak{g}^{*})$ which satisfies (\ref{5.3}). Then the bracket
(\ref{5.10}) coincides with the bracket on $Fun(G^{*})_{e}$  associated
with the central extension of $\frak{g}$.

Remarkably, in formulae (\ref{5.7}, \ref{5.8}, \ref{5.9}, \ref{5.11})
derivations are replaced
by finite automorphisms. This is a manifestation of a general principle:
for quantum groups (and even for Poisson Lie groups) differential
operators are replaced by difference operators. In the present case we
see that deformations of Poisson structures on Lie groups are
parametrized by the group $\mbox{Out}(\frak{g},\frak{g}^{*})$ of outer
automorphisms (which may be considerably larger than the infinitezimal
group $[\mbox{Der}(\frak{g},\frak{g}^{*})]\simeq H^{2}(\frak{g})$).

A typical example when the group $\mbox{Out}(\frak{g},\frak{g}^{*})$ is
non-trivial (although $H^{2}(\frak{g})=0$) is connected with lattice
systems. Let us describe this example in more detail.

Let $(\frak{g},\frak{g}^{*})$ be a factorizable Lie bialgebra. Put
$\cal{G}
{}~=~\oplus^{\mbox{$N$}}\frak{g},~
\cal{G}^{*}~=~\oplus^{\mbox{$N$}}\frak{g}^{*}$. It is convenient
to consider elements of $(\cal G,\cal G^{*})$ as functions on $\bf{Z}
/\mbox{$N$}\bf{Z}$ with values in  $\frak{g},\frak{g}^{*}$. Let
$\tau$ be the
automorphism of $\cal G$ induced by the cyclic permutation on $\bf{Z}
/\mbox{$N$}\bf{Z}$. Clearly, $\tau$ is an automorphism of
$(\cal G,\cal G^{*})$.
In the 'continuous limit` the periodic lattice
$\bf{Z}/\mbox{$N$}\bf{Z}$ is
replaced by a circle and the automorphism $\tau$ by the derivation
$\partial_{x}$  giving rise to the central extension of the loop algebra
$L\frak{g}$. The twisted algebras $Fun(G^{N})_{\tau}$ mimick, in the
finite-dimensional setting, the effects of the central extension of
$L\frak{g}$.

The algebra $Fun(G^{N})$ is generated by the matrix coefficients of the
matrices $\rho_{V}(L^{s}),~s\in\bf{Z}/\mbox{$N$}\bf{Z}$. Specializing
formula
(\ref{5.9}), we get the following Poisson bracket relations
\begin{eqnarray}
\{L_{1}^{k}~,~L_{2}^{k}\}~~=~
{}~-\frac{1}{2}~rL_{1}^{k}L_{2}^{k}~-~\frac{1}{2}~L_{1}^{k}L_{2}^{k}r,
\nonumber\\
\{L_{1}^{k}~,~L_{2}^{k+1}\}~~=~~L_{1}^{k}r_{+}L_{2}^{k+1},
\nonumber\\
\{L_{1}^{k}~,~L_{2}^{l}\}~~=~~0~~~\mbox{for}~~\mid k-l\mid \geq 2.
\label{5.12}
\end{eqnarray}
The bracket (\ref{5.12}) is not {\em ultralocal}, i.e. does not decompose
into direct product of Poisson brackets on different factors. The
properties of this bracket are described by the following
theorem.

\begin{theorem}
\begin{enumerate}
\item[(i)] Let
 $$M:G^{N}\rightarrow
 G:(L^{1},...,L^{N})\mapsto\stackrel{\leftarrow}{\prod}L^{i} $$
 be the monodromy map. Equip the target group $G$ with the Poisson
 bracket (\ref{g*br}). Then $M$ is a Poisson mapping.
 \item[(ii)] Suppose that $N$ is odd. Then the algebra of Casimir
 functions on
 $G^{N}$  is generated by
 \begin{equation}
 c_{k}(L^{1},...,L^{N})~~=~~\rm{tr}~~M^{k},~k=1,2,...
 \label{5.13}
 \end{equation}
 and symplectic leaves in $G^{N}$ coincide with the orbits of lattice
 gauge transformations
 \begin{equation}
 G^{N}\times G^{N}\rightarrow
 G^{N}:(g,L)\mapsto(g_{1}L^{1}g_{2}^{-1},...,g_{N}L^{N}g_{1}^{-1}).
 \label{gauge}
 \end{equation}
 \end{enumerate}
\label{Thm5.5}
 \end{theorem}
 {\em Remark.} The property of the monodromy map described in theorem
 (\ref{Thm5.5}) plays an important role in the theory of integrable
 systems.
 In a more general way, suppose that
 $r\in\rm{End}(\frak{g}\oplus\frak{g})$
 is an arbitrary solution of the modified classical Yang---Baxter
 equation (\ref{cybe}) for the square of $\frak{g}$ ,
 $$r~=~\left(
 \begin{array}{cc}
 A & B\\
 B^{t} & D
 \end{array}
 \right ),~A=-A^{t},D=-D^{t}.$$
 Define the Poisson bracket on $G$ by
 \begin{equation}
 \{L_{1}~,~L_{2}\}_{r}~~=~~AL_{1}L_{2}-L_{1}L_{2}D+L_{1}BL_{2}
 -L_{2}B^{t}L_{1}.
 \label{5.14}
 \end{equation}
 This bracket has an obvious twisted lattice counterpart; the
 corresponding Poisson bracket relations are given by
 \begin{eqnarray}
 \{L_{1}^{k}~,~L_{2}^{k}\}_{r,\tau}~~=~
 ~AL_{1}^{k}L_{2}^{k}-L_{1}^{k}L_{2}^{k}D,\nonumber\\
{}~ \{L_{1}^{k}~,~L_{2}^{k+1}\}_{r,\tau}~~
=~~L_{1}^{k}BL_{2}^{k+1},\nonumber\\
 ~\{L_{1}^{k}~,~L_{2}^{l}\}_{r,\tau}~=
 ~0 ~~\mbox{\rm{for}}~~\mid k-l\mid\geq 2.
 \label{5.15}
 \end{eqnarray}

 \begin{proposition}
 Equip $G^{N}$ with the bracket (\ref{5.15}) and the target group with the
 bracket (\ref{5.14}). Then $M:G^{N}\rightarrow G$ is a Poisson map iff
 $r$ satisfies the constraint
 $$A~-~B^{t}~=~D~-~B.$$
 Functions (\ref{5.13}) are in involution with respect to (\ref{5.15}).
 \label{propMonodr}
 \end{proposition}
 Note that all solutions of the modified classical Yang---Baxter
 equation on $\frak{g}\oplus\frak{g}$ may be completely classified
 (cf.\cite{beldr}) ;the r-matrix (\ref{rdouble}) is a special case of such
 solution.

Let us now turn to the definition of the twisted quantum double. We
shall depart from the definition of the Heisenberg double
$\mbox{$\cal H$}(A)$ of
a factorizable Hopf algebra in terms of the generators $X^{V},Y^{V}$ and
relations (\ref{RXY}). Assume that $\tau$ is an automorphism of
$\rm{End}(V)$ such that $(\tau~\otimes~\tau)R^{VV}~=~R^{VV}$. Put
\begin{equation}
R^{\tau}~~=~~(id~\otimes~\tau)R.
\label{5.16}
\end{equation}
We define $\mbox{$\cal H$}_{\tau}(A)$ as a free algebra generated by
the matrix
coefficients of $X,Y\in \cal H_{\tau}~\otimes~\mbox{\rm{End}}V$
satisfying the
following relations
\begin{eqnarray}
X_{2}X_{1}~~=~~R_{+}X_{1}X_{2}R_{-},\nonumber\\
Y_{2}Y_{1}~~=~~R_{+}Y_{1}Y_{2}R_{-},\nonumber\\
Y_{2}X_{1}~~=~~R_{+}X_{1}Y_{2}R_{+}^{\tau}.
\label{5.17}
\end{eqnarray}
[As in Section 3, we assume that $X,Y$ also satisfy the symmetry
relations which characterize classical groups.]

Clearly, the algebra $\mbox{$\cal H$}_{\tau}(A)$ is a quantization of the
Poisson
algebra $Fun(D_{\tau})$.

Let us denote by $Fun_{q}(G)^{\tau},~Fun_{q}(G^{*})^{\tau}$ the algebras
generated by the matrix coefficients of $T^{V},L^{\pm~V}$ and relations
(\ref{RTTrep}) in which the R-matrix $R^{VV}$ is replaced by
$(R^{VV})^{\tau}$. The following construction is the quantum analogue of
the reduction procedure described in proposition (\ref{5.4}).

\begin{theorem}
\begin{enumerate}
\item[(i)] Formulae
\begin{eqnarray}
C_{L}(X)~~=~~T~\dot{\otimes}~X,~~C_{L}(Y)~~=~~T~\dot{\otimes}~Y,
\nonumber\\
C_{R}(X)~~=~~X~\dot{\otimes}~T^{-1},~~C_{R}(Y)~~=~~Y~\dot{\otimes}~T^{-1}
\label{5.18}
\end{eqnarray}

defines on $Fun_{q}(D)_{\tau}$ the structure of a left $Fun_{q}(G)$- and
a right $Fun_{q}(G)^{\tau}$-comodule.
\item[(ii)] The subalgebra of left coinvariants (i.e. of elements $f\in
Fun_{q}(D)_{\tau}$ satisfying $C_{L}f\in 1\otimes Fun_{q}(D)_{\tau}$) is
generated by the matrix coefficients of $L~=~X^{-1}Y$. The commutation
relations for $L$ are given by
\begin{equation}
L_{2}R_{+}L_{1}(R_{+}^{\tau})^{-1}~~=~
{}~R_{-}^{\tau}L_{1}R_{-}^{-1}L_{2}.
\label{RLLtw}
\end{equation}
\item[(iii)] In a similar way, the subalgebra or right coinvariants is
generated by the matrix coefficients of $M~=~X(Y^{\tau})^{-1}$; the
commutation relations for $M$  are given by
\begin{equation}
R_{+}^{-1}M_{2}R_{+}^{\tau}~~=~
{}~M_{1}(R_{-}^{\tau})^{-1}M_{2}R_{-}.
\label{RMMtw}
\end{equation}
\item[(iv)] Moreover, $L_{1}M_{2}~=~M_{2}L_{1}$, i.e. the subalgebras
of left
and right coinvariants centralize each other in $Fun_{q}(D)_{\tau}$.
\end{enumerate}
\label{mainthm}
\end{theorem}
One can also define another system of generators for
$Fun_{q}(D)_{\tau}$.
\begin{proposition}
The algebra generated by the matrix coefficients of $L^{\pm},T$
satisfying the relations
\begin{eqnarray}
L_{2}^{\pm}L_{1}^{\pm}~~=~~R_{+}L_{1}^{\pm}L_{2}^{\pm}R_{+}^{-1},
\nonumber\\
L_{2}^{-}L_{1}^{+}~~=~~R_{+}L_{1}^{+}L_{2}^{-}(R_{+}^{\tau})^{-1},
\nonumber\\
T_{2}T_{1}~~=~~R_{+}T_{1}T_{2}R_{+}^{-1},\nonumber\\
L_{1}^{\pm}T_{2}~~=~~T_{2}L_{1}^{\pm}R_{\pm} \label{5.19}
\end{eqnarray}
is isomorphic to $Fun_{q}(D)_{\tau}$. The correspondence between the two
sets of generators is given by $X~=~L^{+}T^{-1},~Y~=~L^{-}(T^{\tau})^{-1}$
where $T^{\tau}~=~(id~\otimes~\tau)T$.
\label{prop5.7}
\end{proposition}

The main example of twisting is again related to lattice systems. Let us
consider again the group $G^{N}$ consisting of functions on a periodic
lattice $\bf Z/\mbox{$N$}\bf Z$  with values in $G$. The quantum algebra
$Fun_{q}(G^{N})$ may be described by generators $T^{i},i\in
\bf{Z}/\mbox{$N$}\bf{Z}$
and relations
\begin{equation}
T_{2}^{i}T_{1}^{i}~~=~~RT_{1}^{i}T_{2}^{i}R^{-1},\\
T_{1}^{i}T_{2}^{j}~~=~~T_{2}^{j}T_{1}^{i}~~~\mbox{\rm{for}}~~i\neq j.
\label{5.20}
\end{equation}
Put
\begin{equation}
R^{ij}~~=~~\left\{\begin{array}{ll}
R,i= j,\\
I,i\neq j
\end{array}
\right.
\label{5.21}
\end{equation}
We may regard $R^{ij}$ as a linear operator acting in
$\stackrel{N}{~\otimes~}V$. Let $\tau$  be the cyclic permutation,
$\tau(i)~=~(i+1)~\rm{mod}~N$. Then $R^{\tau i,\tau j}~=~R^{ij}$, and we
may
use $\tau$ to twist our quantum algebras. We get the following algebras
associated with the lattice $\bf{Z}/\mbox{$N$}\bf{Z}$:
\begin{enumerate}
\item[(i)] The twisted double generated by the matrix coefficients of
$X^{i},Y^{i}$ and relations
\begin{eqnarray}
X_{2}^{j}X_{1}^{i}~~=~~R_{+}^{ij}X_{1}^{i}X_{2}^{j}R_{-}^{ij},\nonumber\\
Y_{2}^{j}Y_{1}^{i}~~=~~R_{+}^{ij}Y_{1}^{i}Y_{2}^{j}R_{-}^{ij},\nonumber\\
Y_{2}^{j}X_{1}^{i}~~=~~R_{+}^{ij}X_{1}^{i}Y_{2}^{j}R_{+}^{i,j+1}.
\label{5.22}
\end{eqnarray}
\item[(ii)] The algebra $Fun_{q}(D/G)_{\tau}\simeq G^{*}_{\tau}$ with
generators $L^{i}$ and relations
\begin{equation}
L_{2}^{j}R_{+}^{ij}L_{1}^{i}(R_{+}^{i,j+1})^{-1}~~=~~
R_{-}^{i-1,j}L_{1}^{i}R_{-}^{ij}L_{2}^{j}.
\label{5.23}
\end{equation}
\end{enumerate}
\newtheorem{Mainth}{Theorem \cite{afs}}
\begin{Mainth}
\begin{enumerate}
\item[(i)] Let
$$M~~=~~\stackrel{\leftarrow}{\prod}L^{i}$$
be the monodromy matrix. Then
\begin{equation}
M_{2}R_{+}M_{1}R_{+}^{-1}~~=~~R_{-}M_{1}R_{-}^{-1}M_{2}.
\label{5.24}
\end{equation}
Thus the monodromy map gives rise to an embedding
$$U_{q}(\frak{g})\simeq Fun_{q}(G^{*})\hookrightarrow
Fun_{q}(D/G)_{\tau}$$
\item[(ii)] Suppose that $N$ is odd. Then the centers of the algebras
$Fun_{q}(D/G)_{\tau}$ and $Fun_{q}(G^{*})\subset Fun_{q}(D/G)_{\tau}$
coincide.
\end{enumerate}
\label{Mainthm}
\end{Mainth}
In view of proposition (\ref{Prop3.12}) the last assertion may be
stated in
the following form:
{\em The center of the algebra $Fun_{q}(D/G)_{\tau}$  coincides with the
subalgebra of gauge invariants.}

The algebra $Fun_{q}(D/G)_{\tau}$ may be regarded as a nontrivial
deformation of $U_{q}(\frak{g})^{~\otimes~ N}$. One can show that in the
'continuous limit` this algebra gives the algebra $U(L\frak{g})_{e}$,
i.e. the quotient algebra obtained from the universal enveloping algebra
of a Kac---Moody algebra by 'setting the central charge equal to
constant`(cf. \cite{afs}).
 
\end{document}